\documentclass[fleqn]{revtex4-2}

\usepackage{amsmath}
\usepackage{graphicx}
\usepackage{verbatim}
\usepackage{hyperref}

\begin{document}

\title{Relativistic corrections of order $m\alpha^6$: singular operators and regularization.}

\author{V.I.~Korobov}
\email{korobov@theor.jinr.ru}
\affiliation{Bogoliubov Laboratory of Theoretical Physics, Joint Institute for Nuclear Research, Dubna 141980, Russia}

\begin{abstract}
The finite operators are derived for the nonrecoil [leading order in the $(m/M)$ expansion] relativistic corrections in hydrogenlike atoms and ions at order $m\alpha^6$ in the two- and three-body formalisms beyond the adiabatic approximation. Singular operators that appear in the derivation are analyzed, and various regularization methods are discussed.
\end{abstract}

\maketitle

\section{Introduction}

Precision spectroscopy of the hydrogen molecular ion isotopes is of great importance for testing fundamental physics. Recent experiments \cite{Alighanbari20,Patra20,Kortunov21} demonstrate unprecedented accuracy of relative precision of few ppt. For such precision of theoretical predictions, it is necessary to calculate high-order QED corrections. The most natural way to obtain these corrections is the nonrelativistic quantum electrodynamics (NRQED) approach~\cite{Caswell1986,Kinoshita1996}, which is a powerful tool for studying QED in weakly bound (low-$Z$) few-body systems. It has been extensively used in the past to solve the two-body problem \cite{Kinoshita1996,Pineda98,Eides} and recently for three-body systems such as the helium atom~\cite{Pachucki98,Yelkhovsky,Pachucki2006,Patkos2016} and hydrogen molecular ions~\cite{Korobov2017,Zhong2018}, as well as for four-body systems such as Li and Be$^+$~\cite{Puchalski2014,Puchalski2015}.

From a theoretical point of view, NRQED is the most suitable platform for such accurate calculations of spectra of bound states, since the nonrelativistic Schr\"odinger equation already determines the dynamics of light atoms and molecules with good accuracy. At the same time, the methods of quantum field theory implemented in NRQED make it possible to construct a rigorous perturbation theory for calculating higher-order corrections in terms of the parameters $v/c\sim Z\alpha$ and $\beta=m/M$, where $v$ is the velocity of electrons (muons) in atoms, $m/M$ is the ratio of masses of a light particle to heavy one \cite{Kinoshita1996}. The NRQED Lagrangian is constructed from the non-relativistic fields $\psi$ for the particles entering the system: scalar fields for scalar particles, Pauli spinors for particles with spin $1/2$, etc. The photon is necessarily relativistic and is defined in the same way as in QED. Contributions from QED that involve relativistic loop momenta, the "hard-scale" contributions, are absorbed as renormalizations in the coupling constants for the various contact interactions \cite{Melnikov02}. It is important to note that by construction NRQED is completely equivalent to quantum electrodynamics in the sense that the bound state energy has the same series expansion in $\alpha$ and $Z\alpha$ in both theories.

So far the relativistic corrections in the nonrecoil limit at $m\alpha^6$ order for the hydrogen molecular ion were studied within the adiabatic approximation \cite{Korobov07,Korobov08,Korobov21}. Here we present a new study which is based on the three-body formalism. We start with the nonrecoil effective Hamiltonian at $m\alpha^6$ order. We then propose a method by which the singular part in the effective Hamiltonian may be separated in a form of two singular operators: $\boldsymbol{\mathcal{E}}^2$ and $V(4\pi\rho)$ (see next section for definitions). We demonstrate that it works for both two-particle and three-particle one-electron systems. We show that the singular terms cancel exactly at orders $m\alpha^6$ and $m\alpha^6(m/M)$. The remaining part is finite and all the finite operators that are needed for numerical calculations are explicitly written out.

Another way to derive a set of finite operators for the case of hydrogen molecular ion, in both the nonrecoil limit and the first recoil correction, was investigated in Ref.~\cite{Zhong2018}. It would be interesting to compare the two approaches when numerical results become available.

We investigate various methods of regularization of the problem: coordinate space cutoff, photon propagator modification, and dimensional regularization. Many useful formulas are given that relate the considered methods. Advantages and disadvantages of a particular choice of regularization are discussed.

Atomic units ($\hbar=m=e=1$) are used throughout this work.

\section{Nonrecoil corrections at $m\alpha^6$ order}

First we consider the case of the hydrogen-like atom. The nonrelativistic Hamiltonian in this case is expressed:
\begin{equation}\label{Hamiltonian}
H_0 = \frac{\mathbf{P}^2}{2M}+\frac{\mathbf{p}_e^2}{2m}
      -\frac{Ze^2}{4\pi r} = \frac{\mathbf{p}^2}{2\mu} - \frac{Z}{r}
      = \frac{\mathbf{p}^2}{2\mu} + V,
\end{equation}
where $\mu$ is the reduced mass, $\mu=mM/(m\!+\!M)$, $\mathbf{r} = \mathbf{r}_e\!-\!\mathbf{R}$, and $\mathbf{p} = \mathbf{p_e} = -\mathbf{P}$. Let $\psi_0(r)$ be some bound state solution of the Schr\"odinger equation
\begin{equation}\label{Schr}
H_0\psi_0=E_0\psi_0.
\end{equation}
We assume, unless otherwise stated, that $\langle A \rangle$ is the expectation value of the operator $A$ for the Schr\"odinger equation solution $\psi_0$.

The nonrecoil effective Hamiltonian at $m\alpha^6$ order has been derived in many works (see Refs.~\cite{Pachucki97,Patkos2016,Korobov20}) and may be written
\begin{equation}\label{nrec_H}
\begin{array}{@{}l}
\displaystyle
H^{(6)} = \frac{p_e^6}{16m^5}
     +\frac{\boldsymbol{\mathcal{E}}^2}{8m^3}
     -\frac{3}{64m^4}\left\{p_e^2,\Delta V\right\}
     +\frac{5}{128m^4}\left\{p_e^4,V\right\}
     -\frac{5}{64m^4}\left(p_e^2Vp_e^2\right),
\end{array}
\end{equation}
where $\boldsymbol{\mathcal{E}}=-\boldsymbol{\nabla}V$ and $\Delta V=4\pi\rho$. Curly brackets denote $\{A,B\}=AB+B^*\!A^*$.

The second-order term which contributes at this order is
\begin{equation}\label{sec-order_H}
E_{B}^{(6)} =
\left\langle
   H_B\,Q (E_0-H_0)^{-1} Q\,H_B
\right\rangle,
\end{equation}
where
\[
H_B = \frac{p_e^4}{8m^3}+\frac{Z}{8m^2}4\pi\delta(\mathbf{r}).
\]
Here and below $Q$ is the projection operator onto the subspace orthogonal to $\psi_0$.

Both contributions, see Eqs.~(\ref{nrec_H}) and (\ref{sec-order_H}), diverge, and some care is required to resolve the problem correctly. In what follows we assume that the potential $V$ is regularized in some or other way and the expectation values of the operators that appear in Eqs.~(\ref{nrec_H}) and (\ref{sec-order_H}) obey the following linearity property:
\begin{equation}\label{linearity}
\langle aH_1+bH_2\rangle = a\langle H_1\rangle+b\langle H_2\rangle.
\end{equation}
This is a natural requirement, and, as a rule, all regularization schemes satisfy it.

\subsection{Separation of singularities from the first-order contribution}

In deriving the formulas in this section we use the following notations:
\begin{equation}\label{pmu-op}
[p^4]_\mu = p_e^4-2\mu(4\pi\rho),
\qquad
[p^2]_\mu = p_e^2+2\mu V.
\end{equation}
These operators are nonsingular in the sense that $[p^2]_\mu\psi_0$ is a nonsingular function and $[p^4]_\mu\psi_0$ has at most $1/r^2$ singularity.  Operator $p_e^2[p^2]_\mu\psi_0$ has at most $1/r$ singularity.

Using the linearity property (\ref{linearity}), we can rewrite the expectation values of the operators that appear in Eq.~(\ref{nrec_H}) as follows:
\begin{equation}
\begin{array}{@{}l}\displaystyle
\left\langle p_e^6 \right\rangle =
   -4\mu^2\left\langle V(4\pi\rho) \right\rangle
   +2\mu\left\langle [p^2]_\mu(4\pi\rho) \right\rangle
   +\left\langle [p^2]_\mu [p^4]_\mu \right\rangle
   -2\mu\left\langle V(p_e^2[p^2]_\mu) \right\rangle
   +4\mu^2\left\langle \mathbf{p}_eV^2\mathbf{p}_e \right\rangle,
\\[3mm]\displaystyle
\left\langle Vp_e^4 \right\rangle =
   2\mu\left\langle V(4\pi\rho) \right\rangle
   +\left\langle V(p_e^2[p^2]_\mu) \right\rangle
   -2\mu\left\langle \mathbf{p}_eV^2\mathbf{p}_e \right\rangle,
\\[3mm]\displaystyle
\left\langle p_e^2Vp_e^2 \right\rangle =
   2\mu\left\langle \boldsymbol{\mathcal{E}}^2 \right\rangle
   +2\mu\left\langle V(4\pi\rho) \right\rangle
   +\left\langle p_e^2V[p^2]_\mu \right\rangle
   -2\mu\left\langle \mathbf{p}_eV^2\mathbf{p}_e \right\rangle,
\\[3mm]\displaystyle
\left\langle p_e^2(4\pi\rho) \right\rangle =
   -2\mu\left\langle V(4\pi\rho) \right\rangle
   +\left\langle [p^2]_\mu(4\pi\rho) \right\rangle.
\end{array}
\end{equation}
Only the terms $\left\langle V(4\pi\rho) \right\rangle$ and $\left\langle \boldsymbol{\mathcal{E}}^2 \right\rangle$ in the above expressions are divergent. The finite terms may be further simplified in the two-body case using the relation $[p^2]_\mu\psi_0=2\mu E_0\psi_0$ as follows
\begin{equation}\label{Eq:finite}
\begin{array}{@{}l}\displaystyle
\left\langle p_e^6 \right\rangle' =
   2\mu E_0\left\langle p_e^4 \right\rangle
   -4\mu^2 E_0\left\langle V p_e^2 \right\rangle
   +4\mu^2\left\langle \mathbf{p}_eV^2\mathbf{p}_e \right\rangle =
   \frac{(Z\mu)^6}{n^3}\left(\frac{32}{3}-\frac{56}{3n^2}+\frac{5}{n^3}\right),
\\[2.5mm]\displaystyle
\left\langle Vp_e^4 \right\rangle' =
   \left\langle p_e^2Vp_e^2 \right\rangle' =
   2\mu E_0\left\langle Vp_e^2 \right\rangle
   -2\mu\left\langle \mathbf{p}_eV^2\mathbf{p}_e \right\rangle =
   -\frac{Z(Z\mu)^5}{n^3}\left(\frac{16}{3}-\frac{16}{3n^2}+\frac{1}{n^3}\right),
\\[2.5mm]\displaystyle
\left\langle p_e^2(4\pi\rho) \right\rangle' =
   2\mu E_0\left\langle 4\pi\rho \right\rangle =
   -\frac{4Z(Z\mu)^5}{n^5}.
\end{array}
\end{equation}

\subsection{Separation of singularities from the second-order contribution}

Let us consider the equation for the first-order perturbation wave function $\psi_1$:
\begin{equation}
(E_0-H_0)\psi_1 = QH_B\psi_0.
\end{equation}
The function $\psi_1$ has a $1/r$ singularity due to the presence of $\delta$-function terms on the right-hand side of the equation in $|H_B\psi_0\rangle$, which can be written out explicitly as
\begin{equation}\label{HPH}
H_B\psi_0 =
   -\frac{1}{m^2}\biggl[
      Z\left(\frac{\mu}{m}-\frac{1}{2}\right)\pi\delta(\mathbf{r})
   \biggr]\psi_0
   +\cdots
\end{equation}

Let us try to separate the singular part of the Breit-Pauli wave function solution $\psi_1$ as follows:
\begin{equation}\label{PBmod}
\psi_1 = U(r)\psi_0 + \tilde{\psi}_1.
\end{equation}
where $U(r)=cV(r)$ absorbs a part proportional to $1/r$, and then $\tilde{\psi}_1$ has a smaller singularity, $\tilde{\psi}_1\sim\ln{r}$ at $r\to0$. Substituting $U$ into the initial Schr\"odinger equation
\[
(E_0-H_0)\left(-\frac{cZ}{r}\right) =
   \frac{2cZ}{\mu}\pi\delta(\mathbf{r})+\dots
\]
one gets
\begin{equation}
c = -\frac{\mu(2\mu\!-\!m)}{4m^3}=-\frac{1}{4m}+\mathcal{O}(m/M)\,,
\end{equation}
and $\tilde{\psi}_1$ is a solution of the equation
\begin{equation}\label{eq:s1b}
(E_0-H_0)\tilde{\psi}_1 =
   Q\Bigl[H_B-c\bigl\{E_0\!-\!H_0,V\bigr\}\Bigr]\psi_0.
\end{equation}

\subsection{Modifications to the effective Hamiltonian resulting from the second-order contributions}

Using the transformation
\begin{equation}\label{HPP_H}
H'_{B}=H_B^{}-c\left\{E_0-H_0,V\right\},
\end{equation}
the second-order contribution can be rewritten
\begin{equation}
\begin{array}{@{}l}\displaystyle
\left\langle
   H_B\,Q (E_0-H_0)^{-1} Q\,H_B
\right\rangle =
   \left\langle
      H'_B\,Q (E_0-H_0)^{-1} Q\,H'_B
   \right\rangle
\\[3mm]\hspace{30mm}\displaystyle
   +c\left\langle VH_B+H_BV \right\rangle
   -2c\left\langle V \right\rangle \left\langle H_B \right\rangle
   -c^2\left\langle V(E_0-H_0)V \right\rangle
\end{array}
\end{equation}
and the last three terms may be recast as new interactions in the modified effective Hamiltonian
\begin{equation}\label{H_mB_H}
H^{(6m)} =
   -\frac{c}{8m^3}\left\{V,p^4\right\}+\frac{c}{8m^2}\left\{V,4\pi\rho\right\}
   -2c \left\langle H_B \right\rangle V
   -c^2 V(E_0\!-\!H_0)V.
\end{equation}
The only new singular term is $V(E_0\!-\!H_0)V$ and it is easy to show that it can be transformed as follows:
\begin{equation}\label{Eq:VHV_H}
\left\langle V(E_0\!-\!H_0)V \right\rangle =
   -\frac{1}{2\mu}\left\langle \boldsymbol{\mathcal{E}}^2 \right\rangle.
\end{equation}

Summing the two effective Hamiltonians (\ref{nrec_H}) and (\ref{H_mB_H}), we see that the singular part cancels out exactly in the first two terms of the expansion in $m/M$, and the sum of all finite operators [Eqs.~(\ref{Eq:finite}) and (\ref{H_mB_H})] is
\begin{equation}
\begin{array}{@{}l}
E^{(6)} =
   \left\langle {H^{(6)}}' \right\rangle
   +\left\langle {H^{(6m)}}' \right\rangle
   +\left\langle H'_B\,Q (E_0-H_0)^{-1} Q\,H'_B \right\rangle
\\[3mm]\displaystyle\hspace{15mm}
 = Z^6\left(-\frac{1}{8n^3}-\frac{3}{8n^4}+\frac{3}{4n^5}-\frac{5}{16n^6}\right)
   +Z^6\frac{m}{M}\left[\frac{59}{24n^3}+\frac{27}{8n^4}-\frac{41}{6n^5}+\frac{5}{2n^6}\right].
\end{array}
\end{equation}
The last line is the well known result (see for example Ref.~\cite{Patkos2016}, Appendix D).

Up to this point we have not used any specific regularization, only the linearity property (\ref{linearity}) for regularized divergent operators. In the next two sections we intend to consider these operators and various ways of regularizing them, paying special attention to the multiple relations connecting singular operators and the interrelationships between different regularizations.

\section{Singular identities}

In this section we want to consider in more detail the singular operators appearing in Eqs.~(\ref{nrec_H}) and (\ref{H_mB_H}) and obtain some identities which connect the expectation values of these operators. Again we assume that the Coulomb potential $V=-Z/r$ and its derivatives are regularized.

First we ignore that $\psi_0$ is a solution of the Schr\"odinger equation and will take a more general case when the left- and right-hand side functions inside the brackets are the same function $\psi_0(r)$, which is a smooth function of variable $\mathbf{r}$ (may have a cusp at $r=0$).
Then by using commutation relations and integration by parts one gets
\begin{equation}\label{bi1}
\begin{array}{@{}l}\displaystyle
\left\langle Vp^2V \right\rangle =
   \left\langle V^2p^2 \right\rangle
   -\left\langle V(4\pi\rho) \right\rangle
   +2\left\langle
      V\boldsymbol{\mathcal{E}}\boldsymbol{\nabla}
   \right\rangle,
\\[3mm]\displaystyle
\left\langle Vp^2V \right\rangle =
   \left\langle
      \boldsymbol{\mathcal{E}}^2
   \right\rangle
   -2\left\langle
      V\boldsymbol{\mathcal{E}}\boldsymbol{\nabla}
   \right\rangle
   +\left\langle \mathbf{p}V^2\mathbf{p} \right\rangle,
\\[3mm]\displaystyle
\bigl\langle V(4\pi\rho) \bigr\rangle =
   -\left\langle
      \boldsymbol{\mathcal{E}}^2
   \right\rangle
   +2\left\langle
      V\boldsymbol{\mathcal{E}}\boldsymbol{\nabla}
   \right\rangle.
\end{array}
\end{equation}
The latter may be used as a definition for the distribution $\bigl\langle V(4\pi\rho) \bigr\rangle$. If $\psi_0(\mathbf{r})\to0$, when $r\to0$, it is easy to show that $\bigl\langle V(4\pi\rho) \bigr\rangle=0$, so it is the $\delta$-function type distribution and its support is localized at $r=0$.
Subtracting the last equation from the first one, one gets
\begin{equation}\label{B4}
\left\langle Vp^2V \right\rangle =
   \left\langle
      \boldsymbol{\mathcal{E}}^2
   \right\rangle
   +\left\langle V^2p^2 \right\rangle,
\end{equation}
then summing up second and third equations and taking into account (\ref{B4}):
\begin{equation}\label{B5}
\bigl\langle V(4\pi\rho) \bigr\rangle =
   -\left\langle
      \boldsymbol{\mathcal{E}}^2
   \right\rangle
   -\left\langle V^2p^2 \right\rangle
   +\left\langle \mathbf{p}V^2\mathbf{p} \right\rangle,
\end{equation}
or
\begin{equation}
\left\langle Vp^2V \right\rangle =
   -\bigl\langle V(4\pi\rho) \bigr\rangle
   +\left\langle \mathbf{p}V^2\mathbf{p} \right\rangle.
\end{equation}

If we assume that $\psi_0$ is the solution of the Schr\"odinger equation (\ref{Schr}) we arrive at the identity
\begin{equation}
   \left\langle
     \boldsymbol{\mathcal{E}}^2
   \right\rangle
   -2\mu\left\langle V^3 \right\rangle =
   -\left\langle V(4\pi\rho) \right\rangle
   -2\mu E_0\left\langle V^2 \right\rangle
   +\left\langle\mathbf{p}V^2\mathbf{p}\right\rangle,
\end{equation}
which connects the three basic singular operators.

\section{Regularization}

In this section we assume that $\mu=1$ and all mass scale transformations are ignored. The latter is important for the dimensional regularization. To fully account for the mass dependence in the dimensionally regularized bound-state NRQED we refer to Refs.~\cite{DR-NRQED05,Patkos2016}.

\subsection{Coordinate cutoff regularization}

The divergent integrals can be integrated using coordinate cut-off regularization that can be introduced by setting a lower limit for integration over the radial variable to some small parameter $r_0$.

In a general case one may separate divergent part as follows:
\begin{equation}\label{cut-off1}
\begin{array}{@{}l}\displaystyle
\left\langle V^3 \right\rangle_{\!r_0} =
   -\left\langle
      \frac{Z^3}{r^3}
   \right\rangle_{\!\!r_0} =
   (\ln{r_0}\!+\!\gamma_E)\,Z^3\left\langle4\pi\delta(\mathbf{r})\right\rangle+const
\\[3mm]\displaystyle
\left\langle \boldsymbol{\mathcal{E}}^2 \right\rangle_{r_0} =
   \left\langle
      \frac{Z^2}{r^4}
   \right\rangle_{\!\!r_0} =
   \frac{1}{r_0}\,Z^2\left\langle4\pi\delta(\mathbf{r})\right\rangle
   +(\ln{r_0}\!+\!\gamma_E)\,Z^2\left\langle4\pi\delta'(\mathbf{r})\right\rangle+const
\\[3mm]\displaystyle
\left\langle V\boldsymbol{\mathcal{E}}\boldsymbol{\nabla} \right\rangle_{r_0} =
   \frac{\ln{r_0}\!+\!\gamma_E}{2}\:Z^2\left\langle4\pi\delta'(\mathbf{r})\right\rangle+const
\\[3mm]\displaystyle
\left\langle V(4\pi\rho) \right\rangle_{r_0} \equiv
   -\left\langle
      \boldsymbol{\mathcal{E}}^2
   \right\rangle_{r_0}
   +2\left\langle
      V\boldsymbol{\mathcal{E}}\boldsymbol{\nabla}
   \right\rangle_{r_0} =
   -\frac{Z^2}{r_0}\left\langle 4\pi\delta(\mathbf{r}) \right\rangle
   +Z^2\left\langle 4\pi\delta'(\mathbf{r}) \right\rangle.
\end{array}
\end{equation}
where
\[
\left\langle\phi_1|\delta'(\mathbf{r})|\phi_2\right\rangle =
\left\langle\phi_1\left|
   \frac{\mathbf{r}}{r}\boldsymbol{\nabla}\delta(\mathbf{r})
\right|\phi_2\right\rangle =
   -\left\langle\partial_r\phi_1|\delta(\mathbf{r})|\phi_2\right\rangle
   -\left\langle\phi_1|\delta(\mathbf{r})|\partial_r\phi_2\right\rangle.
\]

Let us introduce the following finite integrals
\begin{equation}\label{Eq:def}
\begin{array}{@{}l}
\displaystyle
\left\langle V^3 \right\rangle_s =
   \lim_{r_0\to0}
   \left\{
   -\left\langle
      \frac{Z^3}{r^3}
   \right\rangle_{\!\!r_0}\!
   -Z^3\left(\ln{r_0}\!+\!\gamma_E\right)\left\langle 4\pi\delta(\mathbf{r}) \right\rangle
   \right\},
\\[4mm]\displaystyle
\left\langle \boldsymbol{\mathcal{E}}^2 \right\rangle_s =
   \lim_{r_0\to0}
   \left\{
   \left\langle
      \frac{Z^2}{r^4}
   \right\rangle_{\!\!r_0}\!
   -Z^2\left[
      \frac{1}{r_0}\left\langle 4\pi\delta(\mathbf{r}) \right\rangle
      +\left(\ln{r_0}\!+\!\gamma_E\right)\left\langle 4\pi\delta'(\mathbf{r}) \right\rangle
   \right]
   \right\},
\\[4mm]\displaystyle
\left\langle V\boldsymbol{\mathcal{E}}\boldsymbol{\nabla} \right\rangle_s =
   \lim_{r_0\to0}
   \left\{
   \left\langle V\boldsymbol{\mathcal{E}}\boldsymbol{\nabla} \right\rangle_{r_0}
   -Z^2\,\frac{\ln{r_0}\!+\!\gamma_E}{2}\left\langle 4\pi\delta'(\mathbf{r}) \right\rangle
   \right\}.
\end{array}
\end{equation}
The first distribution was introduced by Araki \cite{Araki57} and Sucher \cite{Sucher58} for $m\alpha^5$ order corrections for the helium atom.

Calculating divergent operators with derivatives requires a strict definition of how they should be calculated. Otherwise, it leads to ambiguity of results.
For example, the formal expression $\left\langle Vp^2V \right\rangle$ may be written in two ways:
\[
\left\langle V(p^2V) \right\rangle_{r_0} =
   \left\langle
      \boldsymbol{\mathcal{E}}^2
   \right\rangle_{r_0}
   +2\left\langle
      V\boldsymbol{\mathcal{E}}\boldsymbol{\nabla}
   \right\rangle_{r_0},
\]
with no $\delta$-function type terms, and
\[
\left\langle (V\overleftarrow{\mathbf{p}})(\overrightarrow{\mathbf{p}}V) \right\rangle_{r_0} =
   \left\langle
      \boldsymbol{\mathcal{E}}^2
   \right\rangle_{r_0}
   -2\left\langle
      V\boldsymbol{\mathcal{E}}\boldsymbol{\nabla}
   \right\rangle_{r_0}
   +\left\langle \mathbf{p}V^2\mathbf{p} \right\rangle,
\]
where the arrow above $\mathbf{p}$ indicates the direction in which the derivative should act. The two expressions are obviously not equivalent.

\subsection{Mass regularization}

Following the work of Pachucki \cite{Pachucki98,Pachucki97}, the Coulomb photon is modified by introducing a large mass to cut off the high-momentum region:
\begin{equation}
\frac{1}{\mathbf{q}^2}\to
\frac{1}{\mathbf{q}^2}\,\frac{\Lambda^2}{\mathbf{q}^2\!+\!\Lambda^2}.
\end{equation}
Then the Coulomb interaction and its derivatives are expressed as follows:
\begin{equation}
\begin{array}{@{}l}
\displaystyle
V = -\frac{Z}{r} \;\to\; -\frac{Z}{r}\left(1-e^{-\Lambda r}\right),
\\[3mm]\displaystyle
\boldsymbol{\mathcal{E}} = -\boldsymbol{\nabla}V =
   -Z\frac{\mathbf{r}}{r^3} \;\to\;
   -Z\frac{\mathbf{r}}{r^3}
      \left(1-e^{-\Lambda r}-r\Lambda e^{-\Lambda r}\right)
\\[2.5mm]\displaystyle
4\pi\rho = \Delta V = 4\pi Z\delta(\mathbf{r}) \;\to\;
   Z\,\frac{\Lambda^2e^{-\Lambda r}}{r}.
\end{array}
\end{equation}

In a general case of an arbitrary function $\psi_0(r)$ to get a mean value of the regularized operator $A_{\Lambda}$ one may separate integration in three parts:
\begin{equation}\label{reg_int}
\left\langle A_\Lambda\right\rangle =
   \int_0^{r_0} r^2dr\, A_{\Lambda}\,\psi_{\Lambda}^2(r)
   +\int_{r_0} r^2dr\,\left(A_{\Lambda}\,\psi_{\Lambda}^2(r)-A\,\psi^2(r)\right)
   +\int_{r_0} r^2dr\, A\,\psi^2(r).
\end{equation}
where $\psi_\Lambda(r)$ is the solution of the regularized Schr\"odinger equation. Then using the coordinate cut-off regularization, Eq.~(\ref{Eq:def}), one may obtain the following formulas for basic integrals:
\begin{equation}\label{massreg-compare}
\begin{array}{@{}l}\displaystyle
\left\langle V^3_\Lambda\right\rangle =
   -\left\langle \frac{Z^3}{r^3}\right\rangle_{\!s}
   -Z^3 \left\langle 4\pi\delta(\mathbf{r})\right\rangle
      \left(\ln{\Lambda}+\ln{\frac{3}{8}}\right),
\\[3mm]\displaystyle
\left\langle \boldsymbol{\mathcal{E}}_{\Lambda}^2 \right\rangle =
   \left\langle \frac{Z^2}{r^4}\right\rangle_{\!s}
   +\frac{\Lambda}{2}\, Z^2 \left\langle 4\pi\delta(\mathbf{r})\right\rangle
   -Z^2 \left\langle 4\pi\delta'(\mathbf{r})\right\rangle
      \left(\ln{\Lambda}-\ln{2}-\frac{3}{4}\right),
\\[3mm]\displaystyle
\left\langle V_{\Lambda}\boldsymbol{\mathcal{E}}_{\!\Lambda}\boldsymbol{\nabla} \right\rangle =
   \left\langle V\boldsymbol{\mathcal{E}}\boldsymbol{\nabla} \right\rangle_{\!s}
   -\frac{Z^2}{2} \left\langle 4\pi\delta'(\mathbf{r})\right\rangle
      \left(\ln{\Lambda}-\ln{2}-\frac{3}{2}\right).
\end{array}
\end{equation}
Using the identity of Eq.~(\ref{bi1}), one immediately gets
\begin{equation}\label{massreg}
\left\langle V_{\Lambda}(4\pi\rho_{\Lambda}) \right\rangle =
   -\left\langle
      \boldsymbol{\mathcal{E}}^2_{\!\Lambda}
   \right\rangle
   +2\left\langle
      V_{\Lambda}\boldsymbol{\mathcal{E}}_{\!\Lambda}\boldsymbol{\nabla}
   \right\rangle =
   -\frac{Z^2\Lambda}{2}\left\langle4\pi\delta(\mathbf{r})\right\rangle
   +\frac{3Z^2}{4}\left\langle4\pi\delta'(\mathbf{r})\right\rangle.
\end{equation}
The latter expression may be obtained by direct integration of the regularized distribution $V_{\Lambda}(4\pi\rho_{\Lambda})$, which in this case is a regular function.

If $\psi_0$ is a solution of the nonrelativistic Schr\"odinger equation, the expression for $\left\langle \boldsymbol{\mathcal{E}}_{\Lambda}^2 \right\rangle_{\!\Lambda}$ needs some modification. Due to Kato's cusp condition, the equality $\langle 4\pi\delta'(\mathbf{r})\rangle = 2Z\langle 4\pi\delta(\mathbf{r})\rangle$ is satisfied. Then taking into account that
\[
2\left\langle\psi_0\left| \boldsymbol{\mathcal{E}}_{\Lambda}^2 \right|\psi_1\right\rangle =
   2Z^3\left\langle 4\pi\delta(\mathbf{r})\right\rangle\ln{\frac{3}{4}}\,,
\]
where $\psi_1$ is the first order correction to $\psi_0$ in the regularized solution $\psi_{\Lambda}$, we get
\begin{equation}\label{massreg-compare-wf}
\begin{array}{@{}l}\displaystyle
\left\langle \boldsymbol{\mathcal{E}}_{\Lambda}^2 \right\rangle_{\!\Lambda} =
   \left\langle \frac{Z^2}{r^4}\right\rangle_{\!s}
   +2Z^3 \left\langle 4\pi\delta(\mathbf{r})\right\rangle
   \left[\frac{\Lambda}{4Z}
   -\ln{\Lambda}-\ln{\frac{2}{3}}+\frac{3}{4}\right].
\end{array}
\end{equation}
Here $\langle\cdot\rangle_\Lambda$ denotes that the matrix element is calculated using the regularized wave function $\psi_\Lambda$.

We want to stress here that expressions in Eq.~(\ref{massreg-compare}) define functionals on the space of smooth functions with a possible cusp at $r=0$, while Eq.~(\ref{massreg-compare-wf}) determines the functional on the subspace of states satisfying the Kato's cusp condition, which includes as well the solutions of the nonrelativistic Schr\"odinger equation.

\subsection{Dimensional regularization}

As stated in Ref.~\cite{Czarnecki1999} there are two alternative ways to calculate a matrix element in $d$ dimensions. One is to transform it to the coordinate space. A divergence arises at $r\to0$ and in the final result is proportional to the $d$-dimensional $\phi(0)$; the remaining, finite part can be easily calculated in $d=3$.
The alternative approach is to use the $d$-dimensional soltion $\psi_0(p)$ of the Schr\"odinger equation in the momentum space. Both methods lead to the same result.

Let us briefly summarize the main features of the dimensional regularization for bound states following Ref.~\cite{Pachucki2006,Patkos2016}.

The dimensional space is assumed to be $d=3\!-\!2\epsilon$. The Laplacian takes the form
\begin{equation}
\nabla^2=r^{1-d}\partial_rr^{d-1}\partial_r.
\end{equation}

The Coulomb interaction is modified as follows:
\begin{equation}
\begin{array}{@{}l}
\displaystyle
V = -\frac{Z}{r} \;\to\;
   -C_1\frac{Z}{r^{1-2\epsilon}}=
      -\pi^{\epsilon-1/2}\Gamma(1/2-\epsilon)\frac{Z}{r^{1-2\epsilon}},
\\[3mm]\displaystyle
C_1=\frac{\Gamma(1/2-\epsilon)}{\pi^{1/2-\epsilon}}=
   1+\epsilon\left[\gamma_E+\ln{(4\pi)}\right]+\mathcal{O}(\epsilon^2).
\end{array}
\end{equation}
The electric field strength is transformed to
\begin{equation}
\begin{array}{@{}l}
\displaystyle
\boldsymbol{\mathcal{E}} = -\boldsymbol{\nabla}V =
   -Z\frac{\mathbf{r}}{r^3} \;\to\;
   -C_{\mathcal{E}}Z\,\frac{r^i}{r^{3-2\epsilon}}\>,
\\[3mm]\displaystyle
   C_{\mathcal{E}}=(1\!-\!2\epsilon)\,C_1=2\pi^{1-d/2}\Gamma\bigl(d/2\bigr)
   = 1+\epsilon(\gamma_E\!+\!\ln{(4\pi)}\!-\!2)+\mathcal{O}(\epsilon^2).
\end{array}
\end{equation}

The nonrelativistic Hamiltonian $H_\epsilon$ is
\[
H_{\epsilon} = -\frac{\nabla^2}{2}-Z\frac{C_1}{r^{1-2\epsilon}}\,.
\]
Then we look for a solution for small $r$ in the form $\psi_\epsilon(r)=\psi(0)(1-C_Z r^\gamma)$. Substituting it into
the Schr\"odinger equation one gets:
\[
\begin{array}{@{}l}\displaystyle
\gamma = 1+2\epsilon,
\qquad
C_Z = \frac{Z\,C_1}{1\!+\!2\epsilon} =
    \frac{Z\,\Gamma(1/2-\epsilon)}{(1\!+\!2\epsilon)\pi^{1/2-\epsilon}}
    = Z\bigl[1+\epsilon(\gamma_E+\ln{(4\pi)}-2)\bigr]+\mathcal{O}(\epsilon^2).
\end{array}
\]
Thus, at $r\to 0$,
\[
\psi_\epsilon(r) =
   \psi(0)\left(
               1-Z\mu\frac{\Gamma(1/2-\epsilon)}
                      {(1+2\epsilon)\pi^{1/2-\epsilon}}\>r^{1+2\epsilon}
            \right) =
   \psi(0)\Bigl(1-Z\mu\bigl[1+\epsilon(\gamma_E+\ln{(4\pi)}-2)\bigr]r^{1+2\epsilon}\Bigr)+\mathcal{O}(\epsilon^2).
\]

We are now ready to present results relating the dimensionally regularized expectation values of singular operators to the coordinate cut-off ones for the three main distributions:
\begin{equation}
\begin{array}{@{}l}\displaystyle
\left\langle V^3_\epsilon\right\rangle = -Z^3C_1^3\int r^{d-1}dr\,\phi^2(r)\,r^{-3+6\epsilon} =
   -\left\langle \frac{Z^3}{r^3}\right\rangle_{\!s}+Z^3\langle 4\pi\delta(\mathbf{r})\rangle\left(-\frac{1}{4\epsilon}-\frac{3\ln{4\pi}}{4}+\frac{\gamma_E}{4}\right),
\\[3mm]\displaystyle
\left\langle \boldsymbol{\mathcal{E}}_{\epsilon}^2 \right\rangle =
   \left\langle \frac{Z^2}{r^4}\right\rangle_{\!s}
   +Z^2 \left\langle 4\pi\delta'(\mathbf{r})\right\rangle
      \left[-\frac{1}{2\epsilon}-\ln{4\pi}+2\right],
\\[3mm]\displaystyle
\left\langle V_\epsilon\boldsymbol{\mathcal{E}}_{\epsilon}\boldsymbol{\nabla} \right\rangle =
   \left\langle V\boldsymbol{\mathcal{E}}\boldsymbol{\nabla} \right\rangle_{\!s}
   +\frac{Z^2}{2} \left\langle 4\pi\delta'(\mathbf{r})\right\rangle
      \left[-\frac{1}{2\epsilon}-\ln{4\pi}+1\right].
\end{array}
\end{equation}
Using the identity
\[
\bigl\langle V(4\pi\rho) \bigr\rangle =
   -\left\langle
      \boldsymbol{\mathcal{E}}^2
   \right\rangle
   +2\left\langle
      V\boldsymbol{\mathcal{E}}\boldsymbol{\nabla}
   \right\rangle,
\]
one gets
\begin{equation}\label{Vrho-dim}
\left\langle V_{\epsilon}(4\pi\rho_\epsilon)\right\rangle = 0.
\end{equation}
This equality is one of the most important properties of the formalism for the bound-state problem in dimensional regularization. Another way to obtain this identity via $d$-dimensional integration is given in Appendix A, Eq.~(\ref{A3}).

If we use a dimensionally regularized solution of the Schr\"odinger equation, we need to take into account
\[
2\left\langle\psi_0\left| \boldsymbol{\mathcal{E}}_{\epsilon}^2 \right|\psi_1\right\rangle =
   Z^3\left\langle 4\pi\delta(\mathbf{r})\right\rangle\left(\frac{1}{2\epsilon}+\frac{\ln{4\pi}}{2}+\frac{\gamma_E}{2}-1\right),
\]
in a similar way as was done in a previous subsection, and then we come to the identity
\begin{equation}\label{dimreg-compare-wf}
\begin{array}{@{}l}\displaystyle
\left\langle \boldsymbol{\mathcal{E}}_{\epsilon}^2 \right\rangle_{\!\epsilon} =
   \left\langle \frac{Z^2}{r^4}\right\rangle_{\!s}
   +2Z^3 \left\langle 4\pi\delta(\mathbf{r})\right\rangle
   \left[-\frac{1}{4\epsilon}-\frac{3\ln{4\pi}}{4}+\frac{\gamma_E}{4}+\frac{3}{2}\right].
\end{array}
\end{equation}

Expressions for $\left\langle V^3_\epsilon\right\rangle_{\!\epsilon}$ and $\left\langle \boldsymbol{\mathcal{E}}_{\epsilon}^2 \right\rangle_{\!\epsilon}$ obtained here differ from Eqs.~(A22) and (A24) of Ref.~\cite{Patkos2016}. Connection between the two results can be found by introducing some common factor into our regularization scheme:
\[
d^d r\to \exp(\gamma_E\!-\!3\ln{4\pi}\!+\!2)^{2\epsilon} d^d r.
\]
In any case physical quantities should not depend on the choice of the factor, for example, the identity
\begin{equation}
\left\langle \boldsymbol{\mathcal{E}}_{\epsilon}^2 \right\rangle_{\!\epsilon}-2\left\langle V^3_\epsilon\right\rangle_{\!\epsilon} =
 -2E_0\left\langle V^2\right\rangle+\langle \mathbf{p}V^2\mathbf{p} \rangle
\end{equation}
holds for both regularization schemes. The above equation implicitly uses that $\left\langle V_{\epsilon}(4\pi\rho_\epsilon)\right\rangle=0$.

\section{Nonrecoil relativistic corrections at $m\alpha^6$ order for hydrogen molecular ions}

Now let us move on to the case of the molecular hydrogen ion. The nonrelativistic Hamiltonian for the hydrogen molecular ion in the center-of-mass frame (here $\mathbf{r}_a = \mathbf{r}_e\!-\!\mathbf{R}_a$, where $a=1$ and 2) is defined as
\begin{equation}\label{Hamiltonian2}
H_0 = \frac{\mathbf{P}_1^2}{2M_1}+\frac{\mathbf{P}_2^2}{2M_2}+\frac{\mathbf{p}_e^2}{2m}
      -\frac{Z_1}{r_1}-\frac{Z_2}{r_2}+\frac{Z_1Z_2}{R}
 = \frac{\mathbf{p}_1^2}{2\mu_{1}} + \frac{\mathbf{p}_2^2}{2\mu_2}+\frac{\mathbf{p}_1\mathbf{p}_2}{m} + V_1+V_2+V_{12},
\end{equation}
where $\mu_1^{-1}=M_1^{-1}\!+\!m^{-1}$ and $\mu_2^{-1}=M_2^{-1}\!+\!m^{-1}$ are reduced masses of two nuclei, $\mathbf{p}_e=\mathbf{p}_1\!+\!\mathbf{p}_2$, $\mathbf{P}_1=-\mathbf{p}_1$, and $\mathbf{P}_2=-\mathbf{p}_2$.


The nonrecoil effective Hamiltonian and the second-order contribution are expressed in the same way as in the hydrogen case (see Eqs.~(\ref{nrec_H}) and (\ref{sec-order_H}); see as well Ref.~\cite{Korobov20}). The changes required to account for the two nuclei are to replace $V$ with $V_1\!+\!V_2$ in Eq.~(\ref{nrec_H}) and use operator $H_B$ in the following form:
\begin{equation}\label{HB-mol}
H_B = \frac{p_e^4}{8m^3}+\frac{Z_1}{8m^2}4\pi\delta(\mathbf{r}_1)+\frac{Z_2}{8m^2}4\pi\delta(\mathbf{r}_2).
\end{equation}

It should be noted here that in addition to the above contributions, it is necessary to take into account the second-order spin-orbit correction, as was done in the adiabatic approximation \cite{Korobov07,Korobov08}. This term does not contain divergences and is omitted for simplicity from our consideration.

\subsection{Separation of singularities from the first-order contribution}

For further formulas in this section, it will be helpful to introduce the following notations:
\begin{equation}
\begin{array}{@{}l}
V_e = V_1+V_2,
\qquad
\rho_e = \rho_1+\rho_2,
\\[2.5mm]\displaystyle
V_\mu = \mu_1V_1+\mu_2V_2,
\qquad
\rho_\mu = \mu_1\rho_1+\mu_2\rho_2,
\\[2.5mm]\displaystyle
[p^4]_\mu = p_e^4-2(4\pi\rho_\mu),
\qquad
[p^2]_\mu = p_e^2+2V_\mu.
\end{array}
\end{equation}
Again as in the atomic case, the last two operators are nonsingular in the sense that $[p^2]_\mu\psi_0$ is a regular function, $[p^4]_\mu\psi_0$ has at most $1/r^2$ singularity, and $p_e^2[p^2]_\mu\psi_0$ has at most $1/r$ singularity. Here $\psi_0$ is a stationary solution of the Schr\"odinger equation with the nonrelativistic Hamiltonian (\ref{Hamiltonian2}) in the center-of-mass frame.

We use the following identities:
\begin{equation}
\begin{array}{@{}l}\displaystyle
\left\langle V_1(4\pi\rho_e) \right\rangle
   +\left\langle
      \boldsymbol{\mathcal{E}}_1\boldsymbol{\mathcal{E}}_e
   \right\rangle =
   -\left\langle V_1 V_e p_e^2 \right\rangle
   +\left\langle \mathbf{p}_eV_1 V_e\mathbf{p}_e \right\rangle,
\\[3mm]\displaystyle
\left\langle V_1p_e^2V_1 \right\rangle =
   -\left\langle V_1(4\pi\rho_1) \right\rangle
   +\left\langle \mathbf{p}_eV_1^2\mathbf{p}_e \right\rangle,
\\[2mm]\displaystyle
\left\langle V_1p_e^2V_2 \right\rangle =
   -\frac{1}{2}\Bigl[\left\langle V_1(4\pi\rho_2) \right\rangle\!+\!\left\langle V_2(4\pi\rho_1) \right\rangle\Bigr]
   +\left\langle \mathbf{p}_eV_1V_2\mathbf{p}_e \right\rangle,
\end{array}
\end{equation}
which are fulfilled for an arbitrary function $\psi_0$, not only for exact solutions. This allows us to express the operators included in the effective Hamiltonian (\ref{nrec_H}) in the following form:
\begin{equation}
\begin{array}{@{}l}\displaystyle
\left\langle p_e^6 \right\rangle =
   -4\left\langle V_{\mu}(4\pi\rho_\mu) \right\rangle
   +2\left\langle [p^2]_\mu(4\pi\rho_\mu) \right\rangle
   +\left\langle [p^2]_\mu [p^4]_\mu \right\rangle
   -2\left\langle V_{\mu}(p_e^2[p^2]_\mu) \right\rangle
   +4\left\langle \mathbf{p}_eV_\mu^2\mathbf{p}_e \right\rangle,
\\[2.5mm]\displaystyle
\left\langle V_ep_e^4 \right\rangle =
   \Bigl[\left\langle V_e(4\pi\rho_\mu) \right\rangle
      +\left\langle V_\mu(4\pi\rho_e) \right\rangle\Bigl]
   +\left\langle V_e(p_e^2[p^2]_\mu) \right\rangle
   -2\left\langle \mathbf{p}_eV_eV_\mu\mathbf{p}_e \right\rangle,
\\[3mm]\displaystyle
\left\langle p_e^2V_ep_e^2 \right\rangle =
   2\Bigl[\left\langle V_\mu(4\pi\rho_e) \right\rangle
   +\left\langle \boldsymbol{\mathcal{E}}_\mu\boldsymbol{\mathcal{E}}_e \right\rangle\Bigr]
   +\left\langle p_e^2V_e[p^2]_\mu \right\rangle
   -2\left\langle \mathbf{p}_eV_\mu V_e\mathbf{p}_e \right\rangle,
\\[3mm]\displaystyle
\left\langle p_e^2(4\pi\rho_e) \right\rangle =
   -2\left\langle V_\mu(4\pi\rho_e) \right\rangle
   +\left\langle [p^2]_\mu(4\pi\rho_e) \right\rangle.
\end{array}
\end{equation}
Separating the singular part, we arrive at the following set of finite operators:
\begin{equation}\label{Eq:a6finite}
\begin{array}{@{}l}\displaystyle
\left\langle p_e^6 \right\rangle' =
   -4\mu_1\mu_2\bigl\langle V_1(4\pi\rho_2)\!+\!V_2(4\pi\rho_1) \bigr\rangle
   +2\left\langle [p^2]_\mu(4\pi\rho_\mu) \right\rangle
   +\left\langle [p^2]_\mu [p^4]_\mu \right\rangle
   -2\left\langle V_{\mu}(p_e^2[p^2]_\mu) \right\rangle
   +4\left\langle \mathbf{p}_eV_\mu^2\mathbf{p}_e \right\rangle,
\\[2.5mm]\displaystyle
\left\langle V_ep_e^4 \right\rangle' =
   (\mu_1\!+\!\mu_2)\bigl\langle V_1(4\pi\rho_2)\!+\!V_2(4\pi\rho_1) \bigr\rangle
   +\left\langle V_e(p_e^2[p^2]_\mu) \right\rangle
   -2\left\langle \mathbf{p}_eV_eV_\mu\mathbf{p}_e \right\rangle,
\\[3mm]\displaystyle
\left\langle p_e^2V_ep_e^2 \right\rangle' =
   2\bigl\langle \mu_1V_1(4\pi\rho_2)\!+\!\mu_2V_2(4\pi\rho_1) \bigr\rangle
   +2(\mu_1\!+\!\mu_2)\left\langle \boldsymbol{\mathcal{E}}_1\boldsymbol{\mathcal{E}}_2 \right\rangle
   +\left\langle p_e^2(V_e[p^2]_\mu) \right\rangle
   -2\left\langle \mathbf{p}_eV_\mu V_e\mathbf{p}_e \right\rangle,
\\[3mm]\displaystyle
\left\langle p_e^2(4\pi\rho_e) \right\rangle' =
   -2\bigl\langle \mu_1V_1(4\pi\rho_2)\!+\!\mu_2V_2(4\pi\rho_1) \bigr\rangle
   +\left\langle [p^2]_\mu(4\pi\rho_e) \right\rangle.
\end{array}
\end{equation}
and calculation of these operators may be performed directly or indirectly as a product of two nonsingular operators,
\begin{equation}
\langle AB\rangle =
 \sum_{n=1}^N \langle \psi_0| A |\psi_n\rangle\langle \psi_n| B |\psi_0\rangle,
\end{equation}
where $\psi_n$ is a finite set of orthogonal functions, spanning over the finite-dimensional subspace of the Hilbert space of operator $H_0$. Numerical analysis of such computations for the operators presented in Eq.~(\ref{Eq:a6finite}) showed that the convergence is satisfactory and, say, six significant digits in the final result can be obtained at moderate computational time.

\subsection{Separation of singularities from the second-order contributions}

The derivation of the formulas in this part corresponds exactly to what was done in Sec.~II\,B. Here we summarize only those formulas that require modification and are necessary for the final expression.

The transformation of the Breit-Pauli Hamiltonian is expressed as
\begin{equation}\label{HPP}
H'_{B}=H_B^{}-\left\{E_0-H_0,(c_1V_1+c_2V_2)\right\},
\end{equation}
where
\begin{equation}
c_1 = -\frac{\mu_1(2\mu_1\!-\!m)}{4m^3},
\qquad
c_2 = -\frac{\mu_2(2\mu_2\!-\!m)}{4m^3}.
\end{equation}
Matrix elements of $\langle n|H_B'|n'\rangle$ may be obtained directly from Eq.~(\ref{HPP}). The second-order contribution is now transformed:
\begin{equation}
\begin{array}{@{}l}\displaystyle
\left\langle
   H_B\,Q (E_0-H_0)^{-1} Q\,H_B
\right\rangle =
   \left\langle
      H'_B\,Q (E_0-H_0)^{-1} Q\,H'_B
   \right\rangle
\\[3mm]\hspace{30mm}\displaystyle
   +\left\langle V_cH_B+H_BV_c \right\rangle
   -2\left\langle V_c \right\rangle \left\langle H_B \right\rangle
   -\left\langle V_c(E_0-H_0)V_c \right\rangle
\end{array}
\end{equation}
where $V_c=c_1V_1\!+\!c_2V_2$, and the last three terms may be recast as new interactions in the modified effective Hamiltonian
\begin{equation}\label{H_mB}
\begin{array}{@{}l}\displaystyle
H^{(6m)} =
   -\frac{1}{8m^3}\left\{V_c,p_e^4\right\}+\frac{1}{8m^2}\left\{V_c,4\pi\rho_e\right\}
   -2 \left\langle H_B \right\rangle V_c
   -V_c(E_0\!-\!H_0)V_c.
\end{array}
\end{equation}

Again as in the two-body case the new singular terms are $V_1(E_0\!-\!H_0)V_1$ and $V_2(E_0\!-\!H_0)V_2$. The proper treatment of them is considered in Appendix \ref{App.3}, where it is shown that they are transformed to $\left\langle \boldsymbol{\mathcal{E}}^2_a \right\rangle/(2\mu_a)$

The singular terms in the above expressions are $\left\langle V_a(4\pi\rho_a) \right\rangle$ and $\left\langle \boldsymbol{\mathcal{E}}_a^2 \right\rangle$ ($a=1,2$). Summing up the two effective Hamiltonians (\ref{nrec_H}) and (\ref{H_mB}) we see that the singular part cancels out in the first two terms of the expansion in $m/M$; more precisely, the coefficients of the two divergent distributions become exactly zero at the orders $\mathcal{O}(m\alpha^6)$ and $\mathcal{O}(m\alpha^6(m/M))$. The expressions of Eq.~(\ref{Eq:a6finite}) may be used for the free-from-infinities calculation of the $m\alpha^6$ nonrecoil contribution to the spin-averaged energy using the numerical three-body solution of the nonrelativistic Schr\"odinger equation:
\begin{equation}\label{final_3-body}
\begin{array}{@{}l}
E^{(6)} =
   \left\langle {H^{(6)}}' \right\rangle
   +\left\langle {H^{(6m)}}' \right\rangle
   +\left\langle H'_B\,Q (E_0-H_0)^{-1} Q\,H'_B \right\rangle,
\\[3mm]\displaystyle
\left\langle {H^{(6)}}' \right\rangle =
   \frac{\langle p_e^6 \rangle'}{16m^5}
   +\frac{\boldsymbol{\mathcal{E}}_1\boldsymbol{\mathcal{E}}_2}{4m^3}
   -\frac{3}{32m^4}\left\langle p_e^2 (4\pi\rho_e) \right\rangle'
   +\frac{5}{64m^4}\left\langle V_e p_e^4 \right\rangle'
   -\frac{5}{64m^4}\left\langle p_e^2Vp_e^2\right\rangle',
\\[3mm]\displaystyle
\left\langle {H^{(6m)}}' \right\rangle =
   -\frac{1}{4m^3}\left\langle V_c p_e^4 \right\rangle'
   +\frac{1}{4m^2}\left[\langle c_1V_1(4\pi\rho_2)+c_2V_2(4\pi\rho_1)\rangle\right]
   -2 \left\langle H_B \right\rangle \langle V_c \rangle
   -2c_1c_2\left\langle V_1(E_0\!-\!H_0)V_2 \right\rangle,
\end{array}
\end{equation}
where
\begin{equation}
\left\langle V_cp_e^4 \right\rangle' =
   (c_1\mu_2\!+\!c_2\mu_1)\bigl\langle V_1(4\pi\rho_2)\!+\!V_2(4\pi\rho_1) \bigr\rangle
   +\left\langle V_c(p_e^2[p^2]_\mu) \right\rangle
   -2\left\langle \mathbf{p}_eV_cV_\mu\mathbf{p}_e \right\rangle.
\end{equation}

\section{Conclusions}

This work has a twofold interest. First, it shows that all the divergent operators of $m\alpha^6$ order in the nonrecoil approximation are reduced to four operators: $V^3$, $V(4\pi\delta(\mathbf{r}))$, $\boldsymbol{\mathcal{E}}^2$, and $V\boldsymbol{\mathcal{E}\nabla}$. The last two may be used as a definition for the $V(4\pi\delta(\mathbf{r}))$ distribution. In fact, only two of them are needed to express the singular part of the effective Hamiltonians (\ref{nrec_H}) and (\ref{H_mB_H}) and to show that the singular operators are canceled out at the two leading orders $m\alpha^6$ and $m\alpha^6(m/M)$.

The second aim is to analyze the application of different regularizations and to provide a comparison and explicit connection between regularizations, more precisely, to express them through the coordinate cutoff regularization defined in Eq.~(\ref{cut-off1}). It is important to note here that some care is required when considering singular operators, say, $V^4$ and $\boldmath{\mathcal{E}}^2$ are not equivalent distributions in both mass and dimensional regularizations and give different expectation values for the hydrogen bound states $nS$.

Finally, we want to emphasize that only the linearity property (\ref{linearity}) for regularized divergent operators was used to eliminate infinities in both the two-particle and three-particle cases. The case of recoil correction is more complicate and requires careful use of the specific choice of regularization. This is discussed in detail in Ref.~\cite{Korobov25rec}.

\section*{Acknowledgments}

The author would like to express his gratitude to J.-P.~Karr and Zhen-Xiang Zhong for helpful discussions and comments.

\appendix

\section{$V(r)(4\pi\delta(\mathbf{r}))$ distribution in the dimensional regularization.}

The distribution $(1/r)(4\pi\delta(\mathbf{r}))$ in the momentum space is a convolution of two operators:
\begin{equation}
\left(\frac{4\pi}{q_1^2}*\sqrt{\frac{2}{\pi}}\right)(\mathbf{q}) =
   \int d\mathbf{q}_1 \:\mathcal{F}_1(\mathbf{q}_1)\mathcal{F}_2(\mathbf{q}\!-\!\mathbf{q}_1) =
   \int d\mathbf{q}_1 \:\frac{4\pi}{q_1^2}\sqrt{\frac{2}{\pi}}.
\end{equation}
In the dimensional regularization the (massless) integral
\begin{equation}
\int d^d \mathbf{q}\: (\mathbf{q}^2)^{-\alpha}_{} \equiv 0,
\end{equation}
vanishes identically \cite{Collins,Isaev85} for $\alpha\ne d/2$. Thus one obtains
\begin{equation}\label{A3}
\Bigl\langle V(r)(4\pi\delta(\mathbf{r})) \Bigr\rangle \equiv 0.
\end{equation}

\section{Matrix elements for the $nS$ states of the hydrogenlike atom in various regularizations.}

Coordinate cutoff regularized mean values, Eq.~(\ref{Eq:def}), for the $nS$ states of hydrogen are as follows:
\begin{equation}
\begin{array}{@{}l}
\displaystyle
\left\langle V^3 \right\rangle_s =
   \frac{4Z^6}{n^3}
   \left[
      \ln(2Z)+\psi(n)-\psi(1)-\ln{n}
      -\frac{1}{2}+\frac{1}{2n}
   \right],
\\[3mm]\displaystyle
\left\langle \boldsymbol{\mathcal{E}}^2 \right\rangle_s =
   \frac{8Z^6}{n^3}
   \left[
      \ln(2Z)+\psi(n)-\psi(1)-\ln{n}
      -\frac{5}{3}+\frac{1}{2n}+\frac{1}{6n^2}
   \right],
\\[3mm]\displaystyle
\left\langle V\boldsymbol{\mathcal{E}}\boldsymbol{\nabla} \right\rangle_s =
   \frac{4Z^6}{n^3}
   \left[
      \ln(2Z)+\psi(n)-\psi(1)-\ln{n}
      -\frac{2}{3}+\frac{1}{2n}+\frac{1}{6n^2}
   \right],
\\[3mm]\displaystyle
\left\langle V(4\pi\rho) \right\rangle_s \equiv
   -\left\langle \boldsymbol{\mathcal{E}}^2 \right\rangle_s
   +2\left\langle V\boldsymbol{\mathcal{E}}\boldsymbol{\nabla} \right\rangle_s =
   Z^2\left\langle 4\pi\delta'(\mathbf{r}) \right\rangle =
   \frac{8Z^6}{n^3}.
\end{array}
\end{equation}

Mean values for $nS$ states using mass regularization are obtained as follows:
\begin{equation}\label{massreg:nS}
\begin{array}{@{}l}
\displaystyle
\left\langle V_{\Lambda}^3\right\rangle_{\!\Lambda} =
   \frac{4Z^6}{n^3}
   \left[
      \ln{\frac{Z}{\Lambda}}+\ln{\frac{16}{3}}
      +\psi(n)-\psi(1)-\ln{n}
      -\frac{1}{2}+\frac{1}{2n}
   \right],
\\[4mm]\displaystyle
\left\langle \boldsymbol{\mathcal{E}}^2_{\Lambda} \right\rangle_{\!\Lambda} =
   \frac{8Z^6}{n^3}
   \left[
      \frac{\Lambda}{4Z}+\ln{\frac{3Z}{\Lambda}}
      +\psi(n)-\psi(1)-\ln{n}
      -\frac{11}{12}+\frac{1}{2n}+\frac{1}{6n^2}
   \right],
\\[4mm]\displaystyle
\left\langle V_{\Lambda}(4\pi\rho_{\Lambda})\right\rangle_{\!\Lambda} =
   \frac{8Z^6}{n^3}
   \left[
      -\frac{\Lambda}{4Z}-2\ln{\frac{3}{4}}+\frac{3}{4}
   \right].
\end{array}
\end{equation}
Here we use the exact solutions $\psi_{\Lambda}(r)$ for the regularized Schr\"odinger equation in calculations.

Matrix elements for the $d$-dimensional $nS$ states of the hydrogen are expressed as follows:
\begin{equation}
\begin{array}{@{}l}\displaystyle
\left\langle V_{\epsilon}^3 \right\rangle_{\!\epsilon} =
   \frac{4Z^6}{n^3}
   \left[
      -\frac{1}{4\epsilon}
      +\frac{1}{4}\bigl(\gamma_E-3\ln{4\pi}\bigr)
      +\ln{2Z}
      +\psi(n)-\psi(1)-\ln{n}
      -\frac{1}{2}+\frac{1}{2n}
   \right]
\\[3mm]\displaystyle
\left\langle \boldsymbol{\mathcal{E}}_{\epsilon}^2 \right\rangle_{\!\epsilon} =
   \frac{8Z^6}{n^3}
   \left[
      -\frac{1}{4\epsilon}
      +\frac{1}{4}\bigl(\gamma_E-3\ln{4\pi}\bigr)
      +\ln{2Z}
      +\psi(n)-\psi(1)-\ln{n}
      -\frac{1}{6}+\frac{1}{2n}+\frac{1}{6n^2}
   \right]
\\[4mm]\displaystyle
\left\langle V_{\epsilon}(4\pi\rho_\epsilon)\right\rangle_{\!\epsilon} = 0.
\end{array}
\end{equation}

\section{Operators $V_i(E_0-H_0)V_j$}\label{App.3}

In this section we use molecular notation for the coordinates [see Eq.~(\ref{Hamiltonian2})].

\subsection{Operator $V_1(E_0-H_0)V_1$}

Let us introduce the Jacoby coordinates:
\begin{equation}
\left\{
\begin{array}{@{\;}l}
\mathbf{x}_1 = \mathbf{r}_1,
\\[2mm]
\mathbf{y}_1 = \mathbf{r}_2-\frac{M_1}{m\!+\!M_1}\mathbf{r}_1,
\end{array}
\right.
\qquad
\left\{
\begin{array}{@{\;}l}
\mathbf{p}_{x_1} = \frac{M_1}{m\!+\!M_1}\mathbf{p}_2+\mathbf{p}_1,
\\[2mm]
\mathbf{p}_{y_1} = \mathbf{p}_2.
\end{array}
\right.
\end{equation}
Then
\begin{equation}
H_0 = \frac{\mathbf{p}_{x_1}^2}{2\mu_1}+\frac{\mathbf{p}_{y_1}^2}{2\nu_1}+V,
\qquad
[V_1,\mathbf{p}_{y_1}]=0,
\qquad
\left[H_0,\frac{1}{r_1}\right] = \frac{1}{2\mu_1}\left[4\pi\delta(\mathbf{r}_1)+2\,\frac{\mathbf{r}_1}{r_1^3}\boldsymbol{\nabla}_{x_1}\right],
\end{equation}
where $\nu_1 = (M_1\!+\!m)M_2/(M_1\!+\!M_2\!+\!m)$. Then
\[
\begin{array}{@{}l}\displaystyle
\left\langle V_1(E_0-H_0)V_1 \right\rangle =
   -\frac{1}{2\mu_1}\left\langle V_1\mathbf{p}_{x_1}^2V_1 \right\rangle
   -\frac{1}{2\nu_1}\left\langle V_1\mathbf{p}_{y_1}^2V_1 \right\rangle
   -\left\langle V_1^3 \right\rangle
   -\left\langle V_1^2(V_2+V_{12}) \right\rangle
   +E_0\left\langle V_1^2 \right\rangle
\\[3mm]\displaystyle\hspace{10mm}
 = \frac{1}{2\mu_1}\left\langle V_1(4\pi\rho_1) \right\rangle
   -\left\langle V_1^3 \right\rangle
   -\frac{1}{2\mu_1}\left\langle \mathbf{p}_{x_1}V_1^2\mathbf{p}_{x_1} \right\rangle
   -\frac{1}{2\nu_1}\left\langle \mathbf{p}_{y_1}V_1^2\mathbf{p}_{y_1} \right\rangle
   -\left\langle V_1^2(V_2\!+\!V_{12}) \right\rangle
   +E_0\left\langle V_1^2 \right\rangle.
\end{array}
\]
Using
\[
\left\langle V_1\mathbf{p}_{x_1}^2V_1 \right\rangle =
   \left\langle \boldsymbol{\mathcal{E}}_1^2 \right\rangle
   +\left\langle V_1^2\mathbf{p}_{x_1}^2 \right\rangle,
\qquad
\left\langle V_1\mathbf{p}_{y_1}^2V_1 \right\rangle =
   \left\langle V_1^2\mathbf{p}_{y_1}^2 \right\rangle,
\]
one gets
\begin{equation}\label{VHV}
\left\langle V_1(E_0\!-\!H_0)V_1 \right\rangle =
    -\frac{1}{2\mu_1}\left\langle \boldsymbol{\mathcal{E}}^2_1 \right\rangle.
\end{equation}

Yet another equation, which connects the three singular distributions, should be written as follows:
\begin{equation}
\begin{array}{@{}l}\displaystyle
\frac{1}{2\mu_1}\Bigl(\left\langle \boldsymbol{\mathcal{E}}_1^2 \right\rangle+\left\langle V_1(4\pi\rho_1) \right\rangle\Bigr)
   -\left\langle V_1^3 \right\rangle =
\\[3mm]\displaystyle\hspace{15mm}
 = \frac{1}{2\mu_1}\left\langle \mathbf{p}_{x_1}V_1^2\mathbf{p}_{x_1} \right\rangle
   +\frac{1}{2\nu_1}\left\langle \mathbf{p}_{y_1}V_1^2\mathbf{p}_{y_1} \right\rangle
   +\left\langle V_1^2(V_2+V_{12}) \right\rangle
   -E_0\left\langle V_1^2 \right\rangle.
\end{array}
\end{equation}

\subsection{Operator $V_2(E_0-H_0)V_1$}

Using: $\langle V_i(E_0-H_0)V_j \rangle = -\langle V_i[H_0,V_j] \rangle$, one gets
\[
\begin{array}{@{}l}\displaystyle
\left\langle V_2(E_0-H_0)V_1 \right\rangle
 = \frac{1}{2\mu_1}\left\langle V_2 (4\pi\rho_1)\right\rangle
   -\left\langle
      \frac{Z_1Z_2}{\mu_1}\frac{1}{r_{2}r_1^3}\,\bigl(\mathbf{r_1}\boldsymbol{\nabla}_{x_1}\bigr)\right\rangle
 = -\frac{1}{2m}\left\langle \boldsymbol{\mathcal{E}}_1\boldsymbol{\mathcal{E}}_2 \right\rangle.
\end{array}
\]
There are no divergent terms in this case.

\end{document}